\documentstyle[12pt]{article}
\begin{document}
\baselineskip=2\baselineskip
\begin{center}
{\large\bf Slave-Boson Mean-Field Theory of the Antiferromagnetic }  \\
{\large\bf  State in the Doubly Degenerate Hubbard Model }  \\
{\large\bf  - The Half-Filled Case - }  \\
\end{center}

\begin{center}
Hideo Hasegawa$^\dagger$  \\
{\it Department of Physics, Tokyo Gakugei University  \\
Koganei, Tokyo 184, Japan}
\end{center}
\begin{center}
{\rm (February 5, 1997)}
\end{center}
\thispagestyle{myheadings}
%
\begin{center} 
{\bf Abstract}   \par
\end{center} 
 
   The antiferromagnetic ground state of the half-filled
Hubbard model with the doubly degenerate orbital
has been studied by using
the slave-boson mean-field theory which was  previously proposed
by the present author.
Numerical calculations for the simple cubic model have shown
that the metal-insulator transition does not take place
except at the vanishing interaction point,
in strong contrast with  its paramagnetic solution.
The energy gap in the density of states of the 
antiferromagnetic insulator is much reduced by the effect
of electron correlation.
The exchange interaction $J$ plays an important
role in the antiferromagnetism:
although for $J = 0$ the sublattice magnetic moment $m$ 
in our theory is fairly smaller than 
$m_{\rm HFA}$ obtained in the Hartree-Fock approximation, 
$m$ for $J/U > 0.2$ ($U$: the Coulomb interaction) 
is increased to become comparable to $m_{\rm HFA}$. 
Surprisingly, the antiferromagnetic state is easily destroyed if
a small, {\it negative} exchange interaction ($J/U < -0.05$) is
introduced.

\vspace{0.5cm}
\noindent
PACS No. 71.27+a, 71.30.+h, 75.10.Lp
%

%
\newpage
\noindent
\begin{center}
{\large\bf I. INTRODUCTION}
\end{center}

   Much progress has been made in our theoretical understanding
on the effect of electron correlation in systems such as
transition metals and high-$T_c$ materials.
Most of the theoretical studies have been made for the single-band
Hubbard model (SHM)$^{1-3}$ for its simplicity.
Actual systems, however, inevitably have the orbital degeneracy.
It is necessary to investigate the role of the orbital degeneracy
and the effect of Hund-rule coupling due to the exchange 
interaction for a better understanding on strongly 
correlated systems.

  In the last few years the Hubbard model with orbital degeneracy
has been extensively studied by using various methods such as the
Gutzwiller approximation (GA),$^{4-6}$ 
the slave-boson  theory$^{7,8}$, 
the dynamical mean-field approximation,$^{9}$
and the projective self-consistent method.$^{10}$
The original GA proposed by Gutzwiller and Chao$^{11}$ 
was reformulated in Refs.4-6.
In a previous paper$^7$ (referred to as I hereafter), 
the present author developed the slave-boson 
functional-integral method for the
Hubbard model with an arbitrary, orbital degeneracy, by employing 
the method proposed by Dorin and Schlottman$^{12}$ for 
the Anderson lattice model.
Fr\'{e}sard and Kotliar$^{8}$ developed an alternative slave-boson
functional integral method.
These slave-boson theories are the simple generalization of
the Kotliar-Ruckenstein theory for the SHM$^{13}$ 
to that for the degenerate Hubbard model, and their
saddle-point approximation is equivalent to the GA.$^{4-6,11}$
In I we have studied the metal-insulator (MI) transition
of the doubly degenerate Hubbard model (DHM)
in the paramagnetic state. The MI transition
takes place when the interaction strength is 
increased,$^{4,6,9,10}$ just as in the case of the SHM.$^{14}$ 
This MI transition is shown to become
the first-order one in the half-filled case when 
the exchange interaction is included.$^{6,7}$

  We should, however, remind the following facts having been
established for the half-filled SHM:

\noindent
(a) In the paramagnetic (P) state, the MI transition 
is realized for $d = 1, 2, 3$ and $\infty$ in the GA, 
but not for $d = 1, 2$ and 3 in the advanced theory 
going beyond the GA.$^{15-17}$

\noindent
(b) In the antiferromagnetic (AF) state, the MI transition 
occurs in neither $d = 1, 2,$ nor 3$^{18}$ in the advanced theory,
nor in $d = \infty$ even within the GA.$^{19-22}$

\noindent
These facts suggest that it is indispensable to take into account 
the antiferromagnetic state in discussing the MI transition
in DHM.

   One of the advantages of the slave-boson functional integral
method over the GA is that it has the wider applicability than
the GA.  For example, we can deal with the system with
the complicated magnetic structures 
such as the antiferromagnetic state,
by using the Green's function formalism.
We will study in this paper, the antiferromagnetic state of the
DHM by employing our slave-boson mean-field theory,$^7$ 
in order to clarify the above-mentioned 
issue relevant to the MI transition
and the roles of the degeneracy and the exchange interaction.

  The paper is organized as follows:  In the next Sec.II,
we  present a basic formulation of our slave-boson saddle-point 
approximation to deal with the antiferromagnetic state in the DHM,
after briefly reviewing I.
Numerical calculations for the simple-cubic lattice 
are presented in Sec.III.
Section IV is devoted to conclusion and supplementary discussion. 
\vspace{1.0cm}
\noindent
\begin{center}
{\large\bf II. FORMULATION}
\end{center}

   We adopt the Hubbard model with the arbitrary, orbital degeneracy $D$, 
whose Hamiltonian is given by
\begin{equation}
H = \sum_{\sigma} \sum_{i j} \sum_{m m'} t^{m m'}_{ij} 
c^\dagger_{im \sigma} c_{jm' \sigma}
+ \frac{1}{2} \sum_i \sum_{(m,\sigma) \neq (m',\sigma')}
U_{mm'}^{\sigma\sigma'}
c^\dagger_{im \sigma} c_{im \sigma}
c^\dagger_{im' \sigma'} c_{im' \sigma'},
\end{equation}
\noindent
where $c_{im\sigma}$ is an annihilation operator of an electron 
with an orbital index {\it m} 
and spin $\sigma \: (= \uparrow, \downarrow)$ on the lattice site $i$.
The electron hopping is assumed to be allowed 
only between the same sub-band:
$t_{ij}^{mm'} = t_{ij} \delta_{mm'}$, for a simplicity.
The on-site interaction, $U_{mm'}^{\sigma \sigma'}$, is given by
\begin{eqnarray}
U_{mm'}^{\sigma\sigma'} &=& U_0 = U   \;\;\;\;\;\;\;\;\;\;
\mbox{for $m = m', \sigma \neq \sigma'$},  \\
&=& U_1 = U - 2J  \;\; 
\mbox{for $m \neq m', \sigma \neq \sigma'$},  \\
&=& U_2 = U - 3 J  \;\; 
\mbox{for $m \neq m', \sigma = \sigma'$}, 
\end{eqnarray}
where $U$ and $J$ are Coulomb and exchange interactions, respectively.

     In I we employed the boson opertor intoduced by
Dorin and Schlottman,$^{12}$ and used the static approximation 
to get the functional integral representation of the partition
function given by$^7$
\begin{equation}
Z = \int D\xi \int D\nu \int Dm \int Dn 
\int \Pi_{\ell=2}^{2D} Db^{(\ell)} 
\:\: exp \: (- \beta  \Phi),
\end{equation}
with
\begin{equation}
e^{- \beta \Phi} = exp \left( - \beta
\left[ \sum_i 
\sum_m (\xi_{im} m_{im}-\nu_{im} n_{im}) 
+ \Phi_0 \;  \right] \right)  \:
Tr \:\: exp \: (- \beta H_{eff}), 
\end{equation}
\begin{equation}
\Phi_0 = \sum_i \: \sum_{\ell=2}^{2D} \;
\sum_{ m_1 \sigma_1, ...., m_{\ell}\sigma_{\ell}} \;\;
\sum_{(m\sigma,m'\sigma')} \;\; U_{mm'}^{\sigma\sigma'} \;\;
b^{(\ell) \dagger}_{i;m_1 \sigma_1, ...., m_{\ell}\sigma_{\ell}} \:
b^{(\ell)}_{i;m_1 \sigma_1, ...., m_{\ell}\sigma_{\ell}}.
\end{equation}
The effective Hamiltonian is given by
\begin{equation}
H_{eff} = \sum_{\sigma} \sum_{i j} \sum_{m } 
z_{im\sigma}^{\dagger} z_{jm\sigma}
t_{ij} \: 
c^\dagger_{im \sigma} c_{jm' \sigma}
+ \sum_\sigma \sum_i \sum_m 
(\nu_{im} - \sigma \xi_{im}) \:
c^\dagger_{im \sigma} c_{im \sigma},
\end{equation}
with
\begin{equation}
z_{im\sigma } = 2 (2 - n_{im})^{-1/2}
n_{im}^{-1/2}
\left[ \surd \overline{e_i \; p_{im\sigma}}  
+ \surd \overline{p_{im\sigma}} \; b_i^{(2)} 
+ \sum^{2D}_{\ell=3}  
\left( {\sl b}_i^{(\ell-1) } \cdot  
{\sl b}_i^{(\ell)} \right)_{m\sigma} \right].
\end{equation}
In Eqs.(5)-(9), 
$D\xi = \Pi_{im} \xi_{im}$ {\it et al.},
$\xi_{im}$ ($\nu_{im}$) is the exchange (charge) field in
the subband $m$ at the site $i$, and
$m_{im}$ ($n_{im}$) is the magnetic moment (electron number).
The forth summation in eq.(7) is performed over a pair of indices
$(m\sigma, m'\sigma')$ with 
$(m\sigma) \neq (m'\sigma')$ in the configuration:
$\{m_1\sigma_1,m_2\sigma_2,...,m_{\ell}\sigma_{\ell} \}$
occipied by $\ell$ electrons.
The introduced boson operator, $b_i^{(\ell)}$: 
\begin{equation}
b_i^{(\ell)} \equiv
b^{(\ell)}_{i;m_1 \sigma_1, ...., m_{\ell}\sigma_{\ell}},
\end{equation}
projects to the configuration of $\ell$ electrons with pairs of
orbital and spin indices $\{m \sigma \}$. 
Its full contraction,
$(b_i^{(\ell)} \cdot b_i^{(\ell)})$,
and partial contraction,
$(b_i^{(\ell)} \cdot b_i^{(\ell)})_{m\sigma}$,
are defined by
\begin{equation}
\left( {\sl b}_i^{(\ell)} \cdot {\sl b}_i^{(\ell)} \right) 
\equiv \sum_{ m_1 \sigma_1, ...., m_{\ell}\sigma_{\ell}}
b^{(\ell)}_{i;m_1 \sigma_1, ...., m_{\ell}\sigma_{\ell}} \:
b^{(\ell)}_{i;m_1 \sigma_1, ...., m_{\ell}\sigma_{\ell}},
\end{equation}
\begin{equation}
({\sl b}_i^{(\ell)} \cdot {\sl b}_i^{(\ell)})_{m_n\sigma_n} 
\equiv \sum_{ m_1 \sigma_1,m_2 \sigma_2, ...., m_{n-1}\sigma_{n-1},
m_{n+1} \sigma_{m+1}1, ...., m_{\ell}\sigma_{\ell}}
b^{(\ell)}_{i;m_1 \sigma_1,m_2 \sigma_2, ...., 
m_{\ell}\sigma_{\ell}} \:
b^{(\ell)}_{i;m_1 \sigma_1,m_2 \sigma_2, ...., m_{\ell}\sigma_{\ell}}.
\end{equation}
The empty state ($e_i$) and
singly occupied one with a $\sigma$-spin
electron ($p_{im\sigma}$) are expressed in terms of
$n_{im}$, $m_{im}$ and $b_i^{(\ell)}$ for $\ell \geq 2$ as$^7$
\begin{equation}
e_i = ({\sl b}_i^{(0)} \cdot {\sl b}_i^{(0)})
= 1 - \sum_{m} n_{im} + 
\sum_{m\sigma}
\sum_{\ell=2}^{2D} \: \:
[(\ell - 1)/\ell] \:\:
({\sl b}_i^{(\ell)} \cdot {\sl b}_i^{(\ell)})_{m\sigma},
\end{equation}
\begin{equation}
p_{im\sigma} = ({\sl b}_i^{(1)} \cdot {\sl b}_i^{(1)})_{m\sigma}
= (n_{im} + \sigma m_{im})/2 - \sum_{\ell=2}^{2D} 
({\sl b}_i^{(\ell)} \cdot {\sl b}_i^{(\ell)})_{m\sigma}.
\end{equation}
The expression for the functional integral given by eqs.(5)-(14)
is a generalization of the single-band model to 
the degenerated-band model,$^{21)}$
and it has  a transparent physical meaning.

    When we apply our slave-boson functional integral method
developed in I to the DHM ($D = 2$),
the functional integral becomes
\begin{equation}
Z = 
\int \! D\xi\int \! Dm\int D\nu\int \! Dn 
\int \! Dd_{0}\int \! Dd_{1}\int \! \Pi_{\sigma} Dd_{\sigma}
\int \! \Pi_{\sigma} Dt_\sigma \int \! Df 
\:\: exp [- \beta ( \Phi_0+\Phi_1+\Phi_2)],
\end{equation}
\begin{equation}
\Phi_0 = \sum_i \left[ 2 U_0 d_{i0} + 2 U_1 d_{i1} 
+ U_2 (d_{i\uparrow} + d_{i\downarrow})
+ 2(U_0 + U_1 + U_2 ) (t_{i\uparrow} + t_{i\downarrow}+ f_i) \right],
\end{equation}
with
\begin{equation}
\Phi_1 = \sum_{im} \left[ \xi_{im} m_{im} 
+ (\mu - \nu_{im}) n_{im} \right],
\end{equation}
\begin{equation}
\Phi_2 = \int d\varepsilon \; f(\varepsilon)\;  (- 1/\pi) \; Im \:\:
Tr \:\: ln \:  G(\varepsilon). 
\end{equation}
In Eqs.(15)-(18), $f(\varepsilon)$ is the 
Fermi-distribution function, and 
$d_i$, $t_i$ and $f_i$ denote the states with double, 
triple and quadruple occupations, respectively.
In particular for doubly occupied states,
we take into account the three kinds of configurations:
$d_{i0}$ for a pair of electrons on the same orbital 
with opposite spin, $d_{i1}$ on the different  orbital 
with opposite spin,  and $d_{i\sigma}$ on the different  
orbital with same spin $\sigma$. 
%
%
%

The one-particle Green function, $G(\varepsilon)$, in Eq.(18) 
is expressed by
\begin{equation}
G(\varepsilon) = (\varepsilon - H_{eff})^{-1},
\end{equation}
where the effective Hamiltonian, $H_{eff}$, is given by
\begin{equation}
H_{eff} = \sum_{\sigma} \sum_{i j} \sum_{m} 
q_{m\sigma}^{ij}
t_{ij} 
c^\dagger_{im \sigma} c_{jm \sigma}
+ \sum_\sigma \sum_i \sum_m 
(\nu_{im} - \sigma \xi_{im}) \:
c^\dagger_{im \sigma} c_{im \sigma},
\end{equation}
the band-narrowing factor, $q_{m\sigma}^{ij},$ being given by
\begin{equation}
q_{m\sigma}^{ij} = z_{im\sigma} \; z_{jm\sigma},
\end{equation}
with
\begin{eqnarray}
z_{im\sigma}
= \frac{ 2 \left[ 
\surd \overline{p_{i\sigma}} 
( \surd \overline{e_{i}}+\surd \overline{d_{i\sigma}} )
+ ( \surd \overline{d_{i0}}+\surd \overline{d_{i1}} )
( \surd \overline{p_{i-\sigma}}+\surd \overline{t_{i\sigma}} )
+ \surd \overline{t_{i-\sigma}} 
( \surd \overline{d_{i-\sigma}}+\surd \overline{f_{i}} ) \right] }
{(n_{im}+\sigma m_{im})^{1/2} \:
(2 - n_{im} - \sigma m_{im})^{1/2}  }.
\end{eqnarray}
%
%
%
\begin{equation}
e_i = 1 - 2 n_{im} + 2d_{i0} + 2 d_{i1} 
+ d_{i\uparrow} + d_{i\downarrow}
+ 4 (t_{i\uparrow} + t_{i\downarrow}) + 3 f_i,
\end{equation}
\begin{equation}
p_{im\sigma} = (n_{im} + \sigma m_{im})/2  
- (d_{i0} + d_{i1} + d_{i\sigma}) 
-  2 \; t_{i\sigma}
-  t_{i-\sigma} -  f_i.
\end{equation}

    In order to discuss the antiferromagnetic (AF) state,
we divide the crystal into two sublattices, A and B.
We assume that for the AF wave vector $Q$, the relation:
$\varepsilon_{k+Q} = - \varepsilon_k$ holds where
$\varepsilon_k$ is the Fourier transform of the transfer integral,
$t_{ij}$.
We take $\xi_{im}$ in Eq.(20) as the staggered field given by
$\xi_{im} = \xi$  ($- \xi$) for $i \in A$  ($i \in B$), 
the exchange fields in the two subbands being assumed to be the same.
The magnitude of $\xi$ will be determined by the variational
condition, as will be shown shortly (Eq.(31)).

Since the effective transfer integral in Eq.(20) is expressed as
a product form: \\
$z_{im\sigma} t_{ij} z_{jm\sigma}$,
we can express the one-electron Green function in terms of 
the locators defined by$^{23)}$
\begin{equation}
X_{im\sigma} = (\varepsilon - \nu_{im} + \sigma \xi_{im})
/r_{im\sigma},
\end{equation}
where $r_{im\sigma} = (z_{im\sigma})^2 
= r_{Am\sigma}$ and $r_{Bm\sigma}$
for $i \in A$ and  $i \in B$, respectively. 
After a simple calculation, we get $\Phi_2$ given by
\begin{equation}
\Phi_2 =  \int d\varepsilon \; f(\varepsilon) \; ( 1/\pi) \;Im \:\:
\sum_{mk\sigma} ln \: \left( q_{m\sigma}^2  
\left[ X_{Am\sigma}(\varepsilon) X_{Bm\sigma}(\varepsilon) 
- \varepsilon_{k}^2 \right] \right), 
\end{equation}
where the band-narrowing factor, $q_{m\sigma}$, is given by
\begin{equation}
q_{m\sigma} = z_{Am\sigma} \; z_{Bm\sigma}  
= z_{m\sigma} \: z_{m-\sigma} 
= \surd \overline{r_{m\sigma} r_{m-\sigma}},
\end{equation}
because $z_{Am\sigma} = z_{Bm-\sigma} = z_{m\sigma}$ 
and $r_{Am\sigma} = r_{Bm-\sigma} = r_{m\sigma}$.

The mean-field free energy is obtained from the 
saddle-point values of the integration variables for which 
the variational conditions yield the following simultaneous equations:
\begin{equation} 
n_{im} = \sum_\sigma n_{im\sigma},
\end{equation}
\begin{equation} 
m_{im} = \sum_\sigma \sigma n_{im\sigma},
\end{equation}
\begin{equation} 
\mu - \nu_{im} + \sum_\sigma R_{im\sigma} 
(\partial r_{im\sigma} /\partial n_{im} ) = 0,
\end{equation}
\begin{equation} 
\xi_{im} + \sum_\sigma R_{im\sigma} 
(\partial r_{im\sigma} /\partial m_{im} ) = 0,
\end{equation}
\begin{equation}
2 U_0 + \sum_{m \sigma} R_{im\sigma} 
(\partial r_{im\sigma}/\partial d_{i0})  = 0,
\end{equation}
\begin{equation}
2 U_1 + \sum_{m \sigma} R_{im\sigma} 
(\partial r_{im\sigma}/\partial d_{i1})  = 0,
\end{equation}
\begin{equation}
U_2 +  \sum_{m \sigma'} R_{im\sigma'} 
(\partial r_{im\sigma'}/\partial d_{i\sigma})  = 0,
\end{equation}
\begin{equation}
2(U_0 + U_1 + U_2) +  \sum_{m \sigma'} R_{im\sigma'} 
(\partial r_{im\sigma'}/\partial t_{i\sigma})  = 0,
\end{equation}
\begin{equation}
2 (U_0 + U_1 + U_2) + \sum_{m \sigma} R_{im\sigma} 
(\partial r_{im\sigma}/\partial f_i)  = 0,
\end{equation}
In Eqs.(28)-(36), $R_{im\sigma}$ and $n_{im\sigma}$ are given by
\begin{equation}
R_{im\sigma} 
= \partial \Phi_2/\partial r_{im\sigma}
= \int d\varepsilon \; f(\varepsilon) \;
\left(-1/\pi \right) \;
Im \left[ \left( \Omega_{m\sigma}/r_{im\sigma} \right) \;
F_0(\Omega_{m\sigma}) \right],
\end{equation}
\begin{equation}
n_{im\sigma} 
= \int d\varepsilon \; f(\varepsilon) \;
\rho_{im\sigma}(\varepsilon),
\end{equation}
where the local densities of states at the site belonging to
A and B sublattices are expressed by
\begin{eqnarray}
\rho_{im\sigma}(\varepsilon) 
&=&  (-1/\pi) \;
Im \; \left[ K_{Am\sigma}(\varepsilon)/r_{Am\sigma} \right] 
\;\;\;\;\;\;\mbox{($i \in A$)}, \nonumber \\ 
&=&  (-1/\pi) \;
Im \; \left[ K_{Bm\sigma}(\varepsilon)/r_{Bm\sigma} \right]
\;\;\;\;\;\;\mbox{($i \in B$)},  
\end{eqnarray}
with
\begin{equation}
K_{An\sigma}(\varepsilon) = \left[ X_{Bm\sigma}(\varepsilon)
/X_{Am\sigma}(\varepsilon) \right]^{1/2}
F_0(\Omega_{m\sigma}),
\end{equation}
\begin{equation}
K_{Bn\sigma}(\varepsilon) = \left[ X_{Am\sigma}(\varepsilon)
/X_{Bm\sigma}(\varepsilon) \right]^{1/2}
F_0(\Omega_{m\sigma}),
\end{equation}
\begin{equation}
\Omega_{m\sigma}(\varepsilon) 
= \left[ X_{Am\sigma}(\varepsilon) \:
X_{Bm\sigma}(\varepsilon) \right]^{1/2},
\end{equation}
\begin{equation}
F_0(\varepsilon) = \int d\varepsilon' \rho_0(\varepsilon')
/(\varepsilon - \varepsilon'),
\end{equation}
$\rho^0(\varepsilon)$ being the unperturbed density of states.$^{24}$
%
\vspace{1.0cm}
\noindent
\begin{center}
{\large\bf III. NUMERICAL CALCULATIONS}
\end{center}

     Numerical calculations have been performed 
for the simple-cubic model with nearest-neighbor hoppings $t$.
Input parameters for our calculations are the non-interacting
density of states,  $\rho_0(\varepsilon)$, 
the Coulomb and exchange interactions, $U$ and $J$, and the
number of electrons per sub-band, $n$, which is unity for the
half-filled case.
We employed the approximate, analytic expression 
for  $\rho_0(\varepsilon)$ of the simple-cubic lattice, 
given by$^{25}$

\begin{eqnarray}
\rho_0(\varepsilon) &=& A \left[ 9 - \omega^2 \right]^{1/2}
- C \left[ 1 - \omega^2 \right]^{1/2}  
\;\;\:\:\:\:\: 
\mbox{ for $\mid \omega \mid \le 1$}, \nonumber \\
&=& A \left[ 9 - \omega^2 \right]^{1/2}
- B \left[ 1 - ( \mid \omega \mid - 2)^2 \right]^{1/2},  
\;\; \mbox{ for $1 <  \mid \omega \mid \le 3$},  \nonumber \\
&=& 0  \;\;\;\;\;\;\;\;\;\;\;\;\;\;\;\;\;
\mbox{for $\mid \omega \mid > 3$},
\end{eqnarray}
where $\omega = \varepsilon/2t$, $A/2t$ = 0.101081, 
$B/2t= 0.128067$ and $C/2t = 0.02$.
The energy and the interactions are hereafter measured in units
of a half of the total band width, $W/2 = 6 \; t = 1$.
The ground-state energy without interactions calculated
by using Eq.(44) is $\varepsilon_0 = -0.3349$, 
which is in good agreement with the exact value of $- 0.3341$.$^{16}$
Since the relations: $e = f$ and $p_{\sigma} = t_{\sigma}$ 
hold for the half-filled case,
we have to self-consistently solve Eqs.(28)-(43) for nine quantities:
$m, \xi, d_0, d_1, d_{\uparrow}, d_{\downarrow},
t_{\uparrow}, t_{\downarrow}$ and $f$, by using the Newton-Rapson
method.  We performed the integrations  given 
by Eqs.(37) and (38) 
with the use of the contour integral along the complex 
energy axis,$^{26}$ in order to reduce the computational time. 

\vspace{0.5cm}
\begin{center}
\noindent
{\bf  A. $J=0$ Case }
\end{center}

     We firstly show the calculated results for the
vanishing exchange interaction ($J = 0$), 
for which $d_0$ and $d_1$ are equivalent. 
Figure 1 shows the sublattice magnetization, $m$, 
as a function of $U$. The antiferromagnetic state is
realized for an infinitesimally small interaction. 
The magnetic moment increases with increasing $U$ 
and asymptotically approaches the saturated value of 
1.0 $\mu_B$ as $U \rightarrow \infty$.
Because of large fluctuations, $m$ in  GA is much reduced
by more than 50 $\%$ than that in the Hartree-Fock 
approximation (HFA) at $U < 1$.

   Figure 2 shows the spin-dependent local densities of states
for $U = 1.0$. They have traces of the van Hove singularity
of the simple-cubic density of states 
and  clear energy gaps characteristic of
the antiferromagnetic insulator. The energy gap
in the GA ($\Delta_{\rm GA} = 0.114$) is much reduced
compared with that in the HFA  ($\Delta_{\rm HFA} = 0.764$).
The $U$ dependence of the energy gap is plotted in Fig.1.
Both $\Delta_{\rm GA}$ and $\Delta_{\rm HFA}$ increase
with increasing $U$ because $\Delta = 2 \:\xi$,
$\xi$ being the staggered exchange field.
The ratio defined by
$a \equiv \Delta_{\rm GA}/\Delta_{\rm HFA} 
= \xi_{\rm GA}/\xi_{\rm HFA}$ 
is unity in the limits of $U \rightarrow 0$ and 
$U \rightarrow \infty$, and it has a broad minimum of 
$a = 0.12$ at $U \sim 0.6$, above which $a$ again increases: 
$a =$ 0.15, 0.29 and 0.54
for $U = 1.0, 1.5$ and 2.0, respectively.

     The band-narrowing factor, which becomes 
$q_{m\uparrow} = q_{m\downarrow} = q$ in our half-filled model,
is shown as a function of $U$ in Fig.3.
In the P state, $q$ monotonously decreases
with increasing $U$ as shown by the dotted curve,
and it vanishes at 
$U = U_c = 12 \varepsilon_0$ = 4.019 where $U_c$ denotes
the critical interaction for the MI transition.$^{4,6,7}$
On the contrary, the $U$ dependence of $q$ in the AF state
is quite different from that in the P state.
When $U$ is increased from the zero value, $q$ of the AF state
gradually departs from that of the P state, and it has
the minimum value of 0.837 at  $U = 1.4$, above which $q$
increases again.
The effect of electron correlation on the band-narrowing 
factor is not considerable although its effect on the
energy gap (or the exchange field) is  significant.

The $U$-dependence of the occupancies is shown in Fig.4.
At $U = 0$ all the occupancies are 0.0625 (= 1/$2^4$).
When $U$ value is increased, only $d_{\uparrow}$ considerably
increases, approaching unity for $U = \infty$:
$d_0 \; (= d_1)$ and $t_{\uparrow}$ have small peaks at
$U \sim 1$ but decrease for larger $U$.

The $U$ dependence of the ground-state energies, $E$,
is shown in Fig.5.
The ground-state energy of the AF state 
($E_{\rm AF}$) calculated by the GA
is not only lower than that of the P state
($E_{\rm P}$) obtained by the GA 
but also lower than that of the AF state
calculated by the HFA. 
The difference: 
$\Delta E = E_{\rm AF}({\rm GA}) - E_{\rm AF}({\rm HFA})$  
expresses the energy gain by including the effect of fluctuations,
and its maximum value is $- 0.056$ at $U = 0.95$.
The HFA for the N\'{e}el state is a good description of the
half-filled DHM in the 
limit of $U = \infty$.
\vspace{0.5cm}
\begin{center}
\noindent
{\bf B. Finite $J$ Case }
\end{center}

     Next we introduce  the exchange interaction, $J$, into
our calculation.
Figure 6 shows the sublattice magnetization as a function of 
$U + J$ for various choices of the ratio: $J/U = 0.1$, 0.2 and 0.3.  
Note that the magnetization  in the HFA is universal when it is plotted
against $U + J$ because its exchange field is given by
$\xi_{\rm HFA} = (1/2)(U + J) m_{\rm HFA}$.
As the value of $J/U$ is increased, the sublattice magnetization is
increased as expected. This fact is more clearly seen in Fig.7, where
the sublattice magnetization and the band-narrowing factor
for $U = 1.0$ are plotted as a function of
$J$.  The sublattice magnetization of the GA, 
particularly near $J = 0$,
is much increased when the $J$ value is increased,
although such an increase in $m$ is  realized also in the HFA result,
but very small.

   The  interaction dependence of the band narrowing-factor is shown in Fig.8.
It was recently pointed out$^{6,7}$ that, when $J$ is finite,
the first-order MI transition is realized in the P state, 
as is shown by dotted curves;
it occurs at $U + J =$ 2.21, 1.95 and 1.83 for $J/U$ = 0.1, 0.2 
and 0.3, respectively.
Our calculation shows that the situation is quite different in AF state:
$q$ decreases only slightly and never vanishes.
The minimum values of $q$ are 0.939, 0.965 and 0.975 for
$J/U =$ 0.1, 0.2 and 0.3, respectively. 

   Figure 9 shows the occupancies as a function of $U + J$ 
in the typical case of $J/U = 0.1$. 
For $J > 0$ the degeneracy between $d_0$ and $d_1$ is removed
and we get $d_0 < d_1$. When the interaction is increased,
only $d_{\uparrow}$ has an appreciable value
at $U + J > 0.5$, as in the case shown in Fig.4.

   Figure 10 shows the ground-state energies, 
$E_{\rm AF}$(GA), $E_{\rm AF}$(HFA)
and $E_{\rm P}$(GA), as a function of the interaction.
We realized that $E_{\rm AF}$(GA) is the lowest among the three for
any $U$ investigated.
The maximum difference of $\Delta E$ is $- 0.029$ at $U$ = 0.7.

\vspace{1.0cm}
\noindent
\begin{center}
{\large\bf IV. CONCLUSION AND DISCUSSION}
\end{center}

     To summarize, we have studied the antiferromagnetic 
ground state in the DHM, 
employing our slave-boson mean-field theory.$^{7}$
Numerical calculations have shown that the MI transition does not take 
place in the antiferromagnetic solution for the half-filled DHM except
at the vanishing interaction point, which arises form a peculiarity
due to the perfect nesting in the model.
This is in contrast with the result 
in its paramagnetic solution,$^{6,7}$
but is the same as the half-filled SHM,$^{21}$
as was discussed in the Introduction.
Except at $U = J = 0$, 
the stable state is  the antiferromagnetic insulator,
whose energy gap is much reduced by electron correlation. 

   It is worth to make a brief comparison between the results
of the DHM and SHM.
Dashed lines in Figs.1, 3, 6 and 8 show
the interaction dependence 
of $m$, $\Delta$ and $q$ of the SHM.$^{21}$
When we compare these results
with the corresponding ones of the DHM,
we notice that both the results are very similar
provided the exchange interaction is not small; 
$J/U > 0.2$. 
When $J$ is small, however,  $m$, $\Delta$  and $q$ in the DHM are
fairly  smaller
than those in the SHM.  Figure 7 shows that the 
exchange interaction effectively  works to increase  
the magnitude of sublattice moment in the DHM.

   In order to more investigate the role of the exchange 
interaction on antiferromagnetism in the DHM, 
we have repeated a numerical calculation for the {\it negative} $J$, 
although $J$ is conventionally taken to be positive.
We notice in Fig.7 that when the negative exchange interaction
is included, the sublattice magnetization for $U = 1.0$ is
considerably  reduced and it disappears for $J/U < -0.05$.
Figure 11 shows the sublattice magnetization and the band
narrowing factor as a function of $U + J$ for $J/U = -0.02$.
When the interaction is increased, the magnetization first
increases at $U + J < 1.2 $ but decreases at larger interaction.
Surprisingly the antiferromagnetic state disappears at $U + J \geq 2.1$,
which is in strong contrast with the HFA result shown 
by the dotted curve.
Figure 12 shows the interaction dependence of the occupancies.
The behavior of the double occupancies in the negative $J$  case is 
rather different from that in the positive $J$ case shown in Fig.9.
When $J$ is negative, the doubly occupied state with the opposite spin 
between the different subbands is less favorable than that 
within the same subband.
Furthermore the {\it triplet state} expressed by 
$d_{\uparrow}$ (or $d_{\downarrow}$) becomes less stable
than the {\it singlet state} expressed by $d_0$ or $d_1$, 
which works to suppress the antiferromagnetism. 
We get $d_0 > d_1 > d_{\uparrow} \: (= d_{\downarrow})$
in the paramagnetic state at $U + J > 2.1$. 
The first-order MI transition occurs at $U + J = 2.68$.
A material with the anti-Hund-rule coupling ($J < 0$)
would show the unusual behavior if it exists.

   The model Hamiltonian adopted in our study (Eq.(1)) is relevant
to systems with partially filled narrow degenerate bands.
A typical example is ${\rm V}_2{\rm O}_3$, which
is an antiferromagnetic insulator (AFI) in the ground state and which
shows the MI transition between  AFI, paramagnetic metal and 
paramagnetic insulator as a function of the temperature, the pressure 
and/or the chemical substitution.
This phase diagram can be qualitatively understood with the SHM.$^{21,27}$ 
The degenerate Hubbard model has much variety than the SHM
because it has an additional, orbital
degree of freedom. It may show the orbital ordering  besides
the spin ordering.
The stability of the different phases in the degenerate band
depends not only on the values of the various interactions included
in the model Hamiltonian but also on the temperature.
It would be interesting to investigate the temperature-interaction
phase diagram of the DHM, by generalizing our approach$^{21}$ 
in which the effects of electron correlation and 
thermal spin fluctuations are properly taken 
into account.

\vspace{1.0cm}
\noindent{\bf Acknowledgment}  \par
\vspace{0.2cm}

This work is partly supported by
a Grant-in-Aid for Scientific Research from the Japanese 
Ministry of Education, Science and Culture.

\newpage
\newpage

\newpage
\noindent{\large\bf  Figure Captions}   \par
\vspace{1.0cm}
\noindent
{\bf Fig. 1} 
The  sublattice magnetization, $m$, and the energy gap,
$\Delta$, as a function of $U$
of the DHM with $J = 0$ in the GA (solid curve) and 
in the HFA (dotted curve),
the results of the SHM being shown by the dashed curve (Ref.21).
\vspace{0.5cm}

\noindent
{\bf Fig. 2} 
The spin-dependent local densities of states 
for $U = 1.0$ and $J = 0.0$:
$\uparrow$-spin (solid curve) and 
$\downarrow$-spin (dashed curve) components in the GA, and
$\uparrow$-spin (dot-dashed curve) and 
$\downarrow$-spin (dottedcurve) components in the HFA.
\vspace{0.5cm}

\noindent
{\bf Fig. 3} 
The band-narrowing factor, $q$, as a function of $U$
of the DHM with $J = 0$ in the AF state (solid curve)
and in the P state (dotted curve),
the result of the SHM in the AF state being shown 
by the dashed curve (Ref.21).
\vspace{0.5cm}

\noindent
{\bf Fig. 4} 
The occupancies  as a function of $U $ with $J/U = 0 $,
the result of $d_{\uparrow} $ divided by  a factor of ten
being plotted by the dot-dashed curve. 
\vspace{0.5cm}

\noindent
{\bf Fig. 5} 
The $U$-dependence of the ground-state energies $E$  
of the AF state with $J = 0$ in the GA (solid curve)
and in the HFA (dashed curve), and their difference:
$\Delta E = E_{\rm AF}(GA) - E_{\rm AF}(HFA)$. 
The result of the P state in the GA is shown 
by the dotted curve.
\vspace{0.5cm}

\noindent
{\bf Fig. 6} 
The  sublattice magnetization, $m$, as a function of $U + J$
of the DHM with various $J/U $ values in the GA (solid curve) and 
in the HFA (dotted curve),
the result of the SHM being shown by the dashed curve (Ref.21).
\vspace{0.5cm}

\noindent
{\bf Fig. 7} 
The sublattice magnetization for $U = 1.0$ as a function of $J$
in the GA (solid curve) and in the HF (dotted curve),
the result of the band narrowing factor being also plotted (chain curve).
\vspace{0.5cm}

\noindent
{\bf Fig. 8} 
The band-narrowing factor, $q$, as a function of $U + J$
of the DHM with various $J/U$ values 
in the AF state (solid curve) and in the P state (dotted curve),
the result of the SHM in the AF state being shown 
by the dashed curve (Ref.21).
\vspace{0.5cm}

\noindent
{\bf Fig. 9} 
The occupancies for $J/U = 0.1$ as a function of $U + J$,
the result of $d_{\uparrow} $ divided by  a factor of ten
being plotted by the chain curve. 
\vspace{0.5cm}

\noindent
{\bf Fig. 10} 
The interaction dependence of the ground-state energies
of the AF state with $J/U = 0.1$ in the GA (solid curve)
and in the HF (dashed curve), and their difference:
$\Delta E = E_{\rm AF}(GA) - E_{\rm AF}(HFA)$. 
The result of the P state
in the GA is shown by the dotted curve.
\vspace{0.5cm}

\noindent
{\bf Fig. 11} 
The  sublattice magnetization, $m$, as a function of $U + J$
with $J/U = -0.02$ in the GA (solid curve) 
and in the HFA (dotted curve),
the band narrowing factor, $q$, being shown by the chain curve.
\vspace{0.5cm}

\noindent
{\bf Fig. 12} 
The occupancies  for $J/U = -0.02 $ as a function of $U + J$.


\begin{thebibliography}{99}
\bibitem{Saurenbach}The equations relevant to the ferromagnetic state
are given if we set 
$X_{Am\sigma} = X_{Bm\sigma} = X_{m\sigma}$,
$K_{Am\sigma}(\varepsilon) = K_{Bm\sigma}(\varepsilon) 
= F_0(X_{m\sigma})$ and $q_{m\sigma} = z^2_{m\sigma}$
in Eqs.(26)-(43).
\end{thebibliography}
\end{document}